\documentclass[journal]{IEEEtran}

\ifCLASSINFOpdf
\else
   \usepackage[dvips]{graphicx}
\fi
\usepackage{url}

\usepackage{graphicx}
\usepackage{amsmath,bm,graphicx}
\usepackage{amsfonts}
\usepackage{multirow,makecell}
\usepackage{algorithmic,algorithm}
\usepackage{hyperref}
\usepackage{enumitem}
\usepackage{adjustbox} 
\usepackage{bm}  
\usepackage{pifont} 
\begin{document}

\title{Is GAN Necessary for Mel-Spectrogram-based Neural Vocoder?}

\author{Hui-Peng Du, Yang~Ai,~\IEEEmembership{Member,~IEEE}, Rui-Chen Zheng, Ye-Xin Lu,~Zhen-Hua~Ling,~\IEEEmembership{Senior Member,~IEEE}
\thanks{This work was funded by the National Nature Science Foundation of China under Grant 62301521, and the Anhui Provincial Natural Science Foundation under Grant 2308085QF200. (Corresponding author: Yang~Ai)}
\thanks{Hui-Peng Du, Yang Ai, Rui-Chen Zheng, Ye-Xin Lu, and Zhen-Hua Ling are with the National Engineering Research Center of Speech and Language Information Processing, University of Science and Technology of China, Hefei, 230027, China (e-mail: redmist@mail.ustc.edu.cn, yangai@ustc.edu.cn, \{zhengruichen, yxlu0102\}@mail.ustc.edu.cn, and zhling@ustc.edu.cn).}}

\maketitle

\begin{abstract}
Recently, mainstream mel-spectrogram-based neural vocoders rely on generative adversarial network (GAN) for high-fidelity speech generation, e.g., HiFi-GAN and BigVGAN. However, the use of GAN restricts training efficiency and model complexity. Therefore, this paper proposes a novel FreeGAN vocoder, aiming to answer the question of whether GAN is necessary for mel-spectrogram-based neural vocoders. The FreeGAN employs an amplitude-phase serial prediction framework, eliminating the need for GAN training. It incorporates amplitude prior input, SNAKE-ConvNeXt v2 backbone and frequency-weighted anti-wrapping phase loss to compensate for the performance loss caused by the absence of GAN. Experimental results confirm that the speech quality of FreeGAN is comparable to that of advanced GAN-based vocoders, while significantly improving training efficiency and complexity. Other explicit-phase-prediction-based neural vocoders can also work without GAN, leveraging our proposed methods.
\end{abstract}


\IEEEpeerreviewmaketitle

\section{Introduction}

\IEEEPARstart{R}{ecently}, mel-spectrogram-based neural vocoders have played a crucial role in various speech generation tasks, such as speech synthesis \cite{shen2018natural, du2024cosyvoice}, speech enhancement \cite{lu2023mp,liu2024audiosr, lu2024towards}, and speech coding \cite{langman2024spectral,stahl2024bitrate}. 
These methods commonly rely on mel-spectrograms as intermediate representations, requiring a vocoder to reconstruct the speech waveform from the mel-spectrogram. 
Therefore, the performance of the mel-spectrogram-based neural vocoders has a significant impact on the overall quality of the speech generation task.

Early neural vocoders, such as WaveNet \cite{tamamori2017speaker} and SampleRNN \cite{ai2018samplernn}, could not achieve parallel waveform generation and instead relied on an autoregressive generation mode, resulting in extremely low generation efficiency. 
HiNet \cite{ai2020neural} accelerates generation by predicting amplitude and phase spectra in sequence. 
However, it requires extra F0 input and two-stage training, which adds operation complexity.
As a result, flow-based \cite{prenger2019waveglow,ping2020waveflow} and generative adversarial network (GAN)-based \cite{kumar2019melgan,kong2020hifi,ai2023apnet,lee2022bigvgan,kaneko2022istftnet, yamamoto2020parallel}  neural vocoders have gradually emerged. 
While both methods enable parallel waveform generation and significantly improve generation efficiency, the former still suffers from issues such as high model complexity. Although diffusion-based vocoders \cite{nguyen2024fregrad} can avoid using generative adversarial strategies, their inference speed on CPUs is relatively slow due to the complexity of diffusion algorithms, making real-time speech generation impossible.
GAN-based neural vocoders are currently the most popular and widely used, as they typically incorporate additional discriminators for adversarial training alongside the neural vocoder. 

However, the adversarial training strategy of GAN-based neural vocoders significantly increases training complexity. 
In addition, when we need to fine-tune a pre-trained vocoder to adapt to the generation of new data, the parameters of the discriminators also need to be stored, thereby increasing storage costs.
For instance, in HiFi-GAN \cite{kong2020hifi}, the discriminators alone contain nearly five times as many parameters as the generator. 

This leads us to question: \textit{Is GAN truly necessary for mel-spectrogram-based neural vocoders? Can we generate high-quality speech waveform without using adversarial training strategy?} 
Both the mel-spectrogram loss and GAN-based loss are commonly used as training objectives in GAN-based neural vocoders. 
However, the mel spectrogram loss lacks phase information, which is essential for waveform generation \cite{espic2017direct, loweimi2021speech}. 
Thus, the supervision provided by discriminators implicitly includes phase information, motivating us to enhance phase quality in the absence of GAN.

Inspired by the above analysis, this paper proposes a novel mel-spectrogram-based neural vocoder, FreeGAN, fully ``free" from GAN during training. 
FreeGAN adopts a framework that first predicts the amplitude, then predicts the phase, and finally reconstructs the waveform via inverse short-time Fourier transform (iSTFT).
To compensate for the absence of discriminator supervision, we introduce three key strategies in FreeGAN. 
1) \textit{Input Design}: We construct an amplitude prior derived from the mel-spectrogram as the model input, instead of directly using the mel-spectrogram, thereby reducing the difficulty of predicting both the amplitude and phase spectra. 
2) \textit{Network Architecture}: We propose a novel SNAKE-ConvNeXt v2 backbone for enhanced modeling capability. 
We use one block for amplitude prediction and four blocks for phase prediction, emphasizing more accurate phase prediction. 
3) \textit{Loss Function}: We introduce a novel frequency-weighted anti-wrapping (FWAW) phase loss to better capture high-frequency phase components. 
\begin{figure*}
    \centering
    \includegraphics[width=0.87\linewidth]{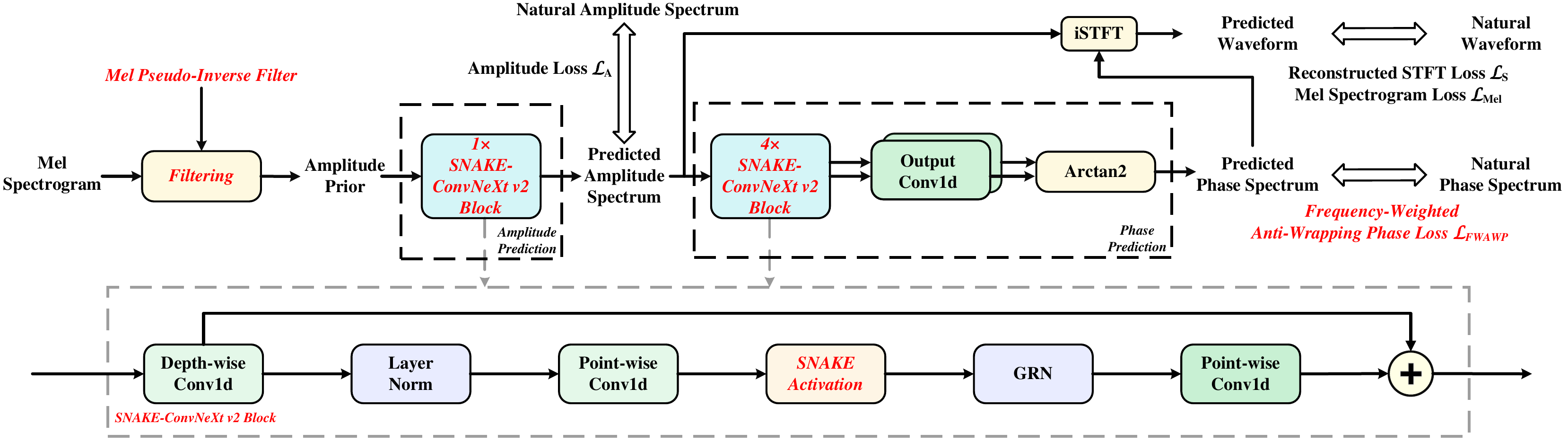}
    \caption{Overview of the proposed FreeGAN vocoder and the strategies we proposed. The strategies we introduced are highlighted in red italics, including the use of pseudo-amplitude spectrum as a prior, SNAKE-ConvNeXt v2 blocks, and frequency-weighted anti-wrapping phase loss. }
    \label{fig:1}
\end{figure*}
Experimental results confirm that FreeGAN performs well without the need for GAN, achieving speech quality comparable to advanced GAN-based vocoders, while offering a simplified training process and lower storage requirements. 
Additionally, we have found that explicit-phase-prediction-based neural vocoders, such as APNet2 \cite{du2023apnet2} and FreeV \cite{lv2024freev}, can also integrate our strategies, thus eliminating the dependence on GAN.

\vspace{-1mm}
\section{Proposed Method}
An overview of the FreeGAN vocoder is illustrated in Figure \ref{fig:1}. 
Unlike APNet \cite{ai2023apnet} and APNet2 \cite{du2023apnet2}, FreeGAN adopts an amplitude-phase serial prediction framework, where the amplitude spectrum is predicted first from the input, followed by the prediction of the phase spectrum, and finally, the speech waveform is reconstructed through iSTFT. 
FreeGAN proposes improvement strategies at three levels (highlighted in red text in Figure \ref{fig:1}), i.e., input, structure, and loss, to compensate for the performance gap caused by the removal of GAN. 
These strategies are explained in detail below.

\vspace{-1mm}
\vspace{-1mm}
\subsection{Input Amplitude Prior}
\vspace{-1mm}
In our previous work \cite{ai2023neural}, the phase spectrum can be accurately predicted from the amplitude spectrum without the use of GAN. 
Therefore, in the serial prediction framework of FreeGAN, phase prediction relies on a higher-quality amplitude spectrum as input.
Inspired by \cite{lv2024freev}, we construct an amplitude prior $\bm{A}^+\in\mathbb R^{F\times N}$ by filtering the mel-spectrogram $\bm{X}\in\mathbb R^{F\times K}$ using the Mel pseudo-inverse filter  $\bm{M}^+\in\mathbb R^{K\times N}$, where $F$ denotes the number of frames, and $N$ and $K$ respectively represent the frequency bins of amplitude spectrum and mel spectogram. 
To ensure the non-negativity of the amplitude spectrum and maintain the stability of training, we use the absolute value of the filtering result and set a small lower bound, i.e.,
\begin{equation}
    \bm{A}^+=\max (|\bm{X}\bm{M}^+|, \epsilon),
\end{equation}
where $\epsilon$ is a small number. 
In FreeGAN, we use the amplitude prior $\bm{A}^+$ as input, instead of the mel spectrogram $\bm{M}$ to predict the amplitude spectrum. . 
Leveraging this prior, we expect to achieve accurate amplitude prediction with a lightweight model, which not only helps reduce model complexity but also contributes to improving subsequent phase prediction accuracy.
\begin{table*}[h!]
	\centering
	\caption{Objective evaluation results and subjective MOS results with 95\% confidence intervals of the proposed FreeGAN vocoder and baselines on the VCTK test set. The `Param' column lists the number of parameters in the vocoder model + the discriminator.
}\label{tab1}
	\adjustbox{width=0.84\textwidth}{
		\renewcommand{\arraystretch}{0.97}
		\begin{tabular}{l c c c c c c c c c}
			\hline
			\hline
    \multirow{2}{*} &{ SNR} & { MCD}& { F0-RMSE} &V/UV error&UTMOS&MOS&GFLOPS&Training Time&Param.\\ &  (dB){ $\uparrow$ }&(dB)$\downarrow$&{ {(cent)$\downarrow$}}&(\%)$\downarrow$&$\uparrow$&$\uparrow$&$\downarrow$&(s/e)$\downarrow$&(M)$\downarrow$\\
   
			\hline
			{Natural}& -&-&-&-&4.04&4.19 ($\pm$0.06)&-&-&-\\
			\hline
			HiFi-GAN&4.15&1.58&31.61&3.97&3.93&-&25.65&546&13.0+70.7\\

			{BigVGAN} & 6.42&0.90&21.04&3.22&3.97&\textbf{4.14 ($\pm$0.06)}& 230.51&1937&115.3+41.3\\
			 {iSTFTNet} &4.15&1.87&32.87&4.13&3.93&- & 19.22&527&\textbf{12.3}+70.7\\
   		{Vocos} & 6.05&0.80&25.17&3.47&3.91&- & \textbf{2.70}&348&13.5+41.7\\
   		{APNet2} & 6.56&0.99&17.38&2.88&\textbf{4.00}&4.13 ($\pm$0.06)& 6.30&445&31.4+41.7\\
			{FreeV} & 6.89&0.81&18.84&2.92&3.97&-& 3.65&381&18.2+41.7\\
			\hline
		 {FreeGAN} & \textbf{7.73}&\textbf{0.65}&\textbf{15.19}&\textbf{2.63}&3.92&4.12 ($\pm$0.06)& \textbf{2.70}&\textbf{191}&13.4+\textbf{0}\\

			\hline
			\hline
	\end{tabular}}
\end{table*}

\begin{table}[h!]
	\centering
	\caption{objective evaluation results for ablation studies.
}\label{tab6}
	\adjustbox{width=\linewidth}{
		\renewcommand{\arraystretch}{0.97}
		\begin{tabular}{l c c c c c}
			\hline
			\hline
    \multirow{2}{*} &{ SNR} & { MCD}& { F0-RMSE} &V/UV error&UTMOS\\ 
    &  (dB){ $\uparrow$ }&(dB)$\downarrow$&{ {(cent)$\downarrow$}}&(\%)$\downarrow$&$\uparrow$\\
			\hline
		{FreeGAN} & 7.73&0.65&15.19&2.63&3.92\\
     \hline
		
   	   {FreeGAN w/o Prior} &7.38&1.40&16.19&2.92&3.81\\	
          {FreeGAN w/o SNAKE} &7.83 &0.65&14.35&2.57&3.87\\
   {FreeGAN w/o FWAWP} & 7.59&0.72 &16.04&2.96&3.92\\
			\hline
			\hline
	\end{tabular}}
\end{table}
\vspace{-1mm}
\vspace{-1mm}
\subsection{Model Structure}
\vspace{-1mm}
As shown in Figure \ref{fig:1}, FreeGAN employs a novel SNAKE-ConvNeXt v2 backbone for both amplitude and phase predictions. 
We use only one block to predict the amplitude spectrum from the amplitude prior. 
For phase prediction, we increase model capacity by employing four blocks and a parallel estimation architecture \cite{ai2023neural}, thereby yielding a more accurate phase spectrum. 


The SNAKE-ConvNeXt v2 block is an improvement of the ConvNeXt v2 block \cite{woo2023convnext}, which has been applied in various speech generation tasks \cite{jiang2024mdctcodec, du2024bivocoder} and shown promising results. 
Specifically, we replaced the Gaussian error linear unit (GELU) \cite{hendrycks2016gaussian} activation function in the original ConvNeXt v2 block \cite{woo2023convnext} with the SNAKE activation function \cite{ziyin2020neural} to further enhance its modeling capabilities. 
The SNAKE activation function is defined as 
\begin{equation}
    f_{SNAKE}(x)=x+\frac{1}{\alpha}\sin ^2(\alpha x),
\end{equation}
where 
$\alpha$ is a learnable parameter. 
\vspace{-1mm}
\subsection{Training Criteria}
\vspace{-1mm}
\label{ssec:pp}
Most GAN-based neural vocoders incorporate additional loss functions—beyond the adversarial loss—primarily to learn amplitude information. 
Given this, we placed greater emphasis on phase learning once GAN was removed. 
Our preliminary experiments showed that high-frequency phase prediction degraded noticeably after removing GAN. 
Based on this observation, we design a novel FWAW phase loss. 
The proposed FWAW phase loss extends the original anti-wrapping phase loss from \cite{ai2023neural} by assigning different loss weights across frequency bins. 
We define a weight factor $w=e^{\frac{\ln{\rho}}{N-1}}$ and construct the weight vector $\bm{w} = [w^0,w^1,w^2, \dots, w^{N-1}]^\top \in \mathbb{R} ^N$, where $\rho$ is a hyperparameter. 
Given the predicted phase spectrum $\hat{\bm{P}}\in\mathbb{R}^{F\times N}$ and natural one $\bm{P}\in\mathbb{R}^{F\times N}$, the FWAW phase-related loss is defined as
\begin{equation}
    \mathcal{L}^{*}_{FWAWP}=\frac{1}{FN}\mathbb E_{(\hat{\bm{P}},\bm{P})}\left[ \bm{v}^\top f_{AW}(\Theta \hat{\bm{P}}-\Theta\bm{P})\bm{w}\right],
\end{equation}
where $\bm{v}=[1,1,\dots,1]^\top\in\mathbb R^N$ is an all-one vector, and $f_{AW}(x)=\left| x-2\pi\cdot round\left( \dfrac{x}{2\pi} \right) \right|$ is the anti-wrapping function \cite{ai2023neural}. 
$\Theta$ represents the differential operator. 
For the FWAW instantaneous phase loss $\mathcal{L}_{FWAWP}^{IP}$, $\Theta$ is set to none. For the FWAW group delay loss $\mathcal{L}_{FWAWP}^{GD}$ and FWAW instantaneous angular frequency loss $\mathcal{L}_{FWAWP}^{IAF}$, $\Theta$ is set to frequency difference and time difference, respectively. 
The final FWAW phase loss is defined as
\begin{equation}
\label{eq4}
\mathcal{L}_{FWAWP}=\mathcal{L}_{FWAWP}^{IP}+\mathcal{L}_{FWAWP}^{GD}+\mathcal{L}_{FWAWP}^{IAF}.
\end{equation}

Besides the phase-related loss, we also introduce amplitude loss $\mathcal{L}_{A}$, reconstructed STFT loss $\mathcal{L}_{S}$, and mel spectrogram loss $\mathcal{L}_{Mel}$ used in our previous work \cite{du2023apnet2}, and combine them for training FreeGAN, i.e.,
\begin{equation}
\label{eq4}
    \mathcal{L}=\mathcal{L}_{FWAWP}+\lambda_A\mathcal{L}_{A}+\lambda_{S}\mathcal{L}_{S}+\lambda_{Mel}\mathcal{L}_{Mel},
\end{equation}
where $\lambda_A, \lambda_S$, and $\lambda_{Mel}$ are hyperparameters.
\vspace{-1mm}
\section{Experiments Setup}
\vspace{-1mm}
\subsection{Dataset}
\vspace{-1mm}
We employed the VCTK-0.92 dataset \cite{veaux2016superseded} and downsampled its utterances to 16 kHz. 
The dataset includes speech recordings from 108 English speakers, with a duration of approximately 44 hours. 
Among the data from 100 speakers, we randomly selected 90\% for the training set and the remaining 10\% for the validation set.
Then we chose 2,937 utterances from the remaining unseen 8 speakers to form the test set.
\vspace{-2mm}
\subsection{Implementation}
\vspace{-1mm}
For FreeGAN vocoder\footnote{Examples of generated speech can be found at {https://redmist328.github.io/FreeGAN/}.}, the amplitude and phase spectra were computed using the STFT with a frame length, frame shift, and FFT size of 320, 80, and 1024 (i.e., $N=513$), respectively. 
The mel-spectrogram was extracted with the same configuration, with a dimensionality of 80 (i.e., $K=80$). 
The hyperparameter settings were as follows: $\epsilon=10^{-5}$, $\rho=2.5$, $\lambda_A=0.45$, $\lambda_S=0.2$ and $\lambda_{Mel}=0.45$. 
The configuration of the trainable layers in the model remained consistent with that in APNet2 \cite{du2023apnet2}.
The model was trained using the AdamW optimizer \cite{kingma2014adam} for up to 1 million steps. 
\vspace{-1mm}
\vspace{-1mm}
\vspace{-1mm}
\subsection{Baselines}
\vspace{-1mm}
We compared FreeGAN with HiFi-GAN \cite{kong2020hifi}, BigVGAN \cite{lee2022bigvgan}, iSTFTNet \cite{kaneko2022istftnet}, Vocos \cite{siuzdakvocos}, APNet2 \cite{du2023apnet2}, and FreeV \cite{lv2024freev}.
For fair comparison, We also reproduced their GAN-free versions (denoted by `w/o GAN'), as well as GAN-free versions that incorporate the improvement strategies we proposed (denoted by `* w/o GAN'). 
However, for the latter version, BigVGAN, HiFi-GAN, iSTFTNet, and Vocos cannot incorporate the FWAW phase loss, so this loss was omitted.

\vspace{-1mm}
\vspace{-1mm}
\subsection{Evaluation Metrics}
\vspace{-1mm}
In this study, we employed five objective metrics to assess
the quality of synthesized speech, including signal-to-noise ratio (SNR), mel-cepstrum distortion (MCD), root mean square error of F0 (F0-RMSE), voiced/unvoiced (V/UV) error, and UTMOS \cite{saeki2022utmos}.
Additionally, we measured the computational complexity of each model using floating-point operations (FLOPs) required to generate 1-second speech. 
To compare the training speeds, we utilized an NVIDIA A100-SXM4-80GB GPU and recorded the time taken by each model to complete one training epoch under identical training parameters.

To evaluate subjective quality, we conducted both mean opinion score (MOS) and ABX preference tests. 
Each MOS test included 20 test utterances along with natural samples. 
Feedback was collected from at least 25 native English speakers via the Amazon Mechanical Turk (AMT) crowdsourcing platform. 
Listeners were asked to rate the naturalness of each utterance on a scale from 1 to 5, with a 0.5 point interval. 
For the ABX test, listeners were tasked with determining which utterance in each pair exhibited better speech quality or whether they had no preference for 20 paired test utterances.

\vspace{-1mm}
\vspace{-1mm}
\section{Results and Analysis}
\vspace{-0.4mm}
\subsection{Is GAN Necessary for the FreeGAN vocoder?}
To investigate whether GAN is essential for the FreeGAN vocoder, we compared it against GAN-based baselines.
Both objective and subjective experimental results are presented in Table \ref{tab1}. 
For objective metrics such as SNR, MCD, F0-RMSE, and U/V error, FreeGAN significantly outperformed all other baseline vocoders. 
For objective UTMOS metric, the FreeGAN performed slightly worse than some of the baselines, e.g., BigVGAN, APNet2 and FreeV. 
To further validate these findings, we conducted a mean opinion score (MOS) test, comparing FreeGAN with baselines that have higher UTMOS scores (including BigVGAN and APNet2), as well as natural speech, to evaluate their subjective naturalness. 
Interestingly, MOS results indicate that FreeGAN is comparable to other baseline vocoders (the p-values of \textit{t}-test compared to BigVGAN and APNet2 are 0.61 and 0.76, respectively). 
Therefore, both objective and subjective results demonstrate that FreeGAN can achieve comparable speech quality to other advanced GAN-based neural vocoders, despite not relying on GAN. 
GAN is unnecessary for our proposed
FreeGAN vocoder.

The efficiency and storage evaluation results are shown in Table \ref{tab1}. 
It can be observed that FreeGAN requires the fewest floating-point operations, making it suitable for applications in resource-constrained environments. 
Regarding training time, since FreeGAN does not require training a discriminator, its training time is significantly shorter than that of other vocoders. 
For example, the training speed of FreeGAN is nearly 15 times faster than that of BigVGAN, but their subjective quality is comparable. However, BigVGAN is designed for general-purpose, zero-shot audio generation, which explains its larger model size and higher training cost. Therefore, the comparison on the VCTK dataset mainly reflects FreeGAN's efficiency in a more constrained setting.
Regarding the storage size of the vocoder model, our proposed FreeGAN has only 13.4M parameters, which is comparable to some lightweight vocoders (e.g., HiFi-GAN, iSTFTNet and Vocos). 
However, the discriminator parameters of GAN-based vocoders are quite large, whereas FreeGAN has eliminated this storage overhead. 
Thus, FreeGAN can improve training efficiency and reduce model storage costs while achieving high-quality speech generation.

To validate the effectiveness of  the three proposed improvement strategies, we conducted ablation experiments on FreeGAN. 
Specifically, we tested three variants: replacing the Mel pseudo-inverse filter with a learnable linear layer (`w/o Prior'), employing the original ConvNeXt v2 block \cite{woo2023convnext} (`w/o SNAKE'), and adopting the original anti-warpping phase loss \cite{ai2023neural} (`w/o FWAWP') in FreeGAN. 
Objective evaluation results are listed in Table \ref{tab6}. 
It can be observed that removing the amplitude prior has the greatest impact on performance. 
Although we used a linear layer to learn the filter coefficients, its effectiveness was not as good as the Mel pseudo-inverse filter. 
In particular, MCD and UTMOS dropped significantly, indicating that the proposed amplitude prior played a crucial role in improving amplitude prediction quality, which in turn enhanced the overall speech quality.
The SNAKE activation in SNAKE-ConvNeXt v2 and FWAW phase loss contributed to improvements in overall speech perceptual quality and phase quality, as indicated by UTMOS and SNR, respectively. However, their contribution is slightly smaller compared to the
amplitude prior.

\vspace{-1mm}
\begin{table}[t!]
	\centering
	\caption{UTMOS results of baseline vocoders, their GAN-free versions (`w/o GAN') and  GAN-free versions with our proposed improvement strategies (`* w/o GAN').}\label{tab2}{
		\adjustbox{width=\linewidth}{
				\renewcommand{\arraystretch}{0.97}
			\begin{tabular}{l c c c}
				\hline
				\hline
& Spectrum Modeling  &{ Explicit Phase Prediction}& {UTMOS$\uparrow$}\\ 
				\hline
			{HiFi-GAN}&\ding{55}&\ding{55} &3.93\\
   		{HiFi-GAN w/o GAN}&\ding{55}&\ding{55} &3.71\\
      	{HiFi-GAN* w/o GAN}&\ding{55}&\ding{55} &3.78\\
       \hline
   		{iSTFTNet}&\ding{51}&\ding{55} &3.93\\
   		{iSTFTNet w/o GAN}&\ding{51}&\ding{55} &3.69\\
   		{iSTFTNet* w/o GAN}&\ding{51}&\ding{55} &3.70\\
     \hline
   		{Vocos}&\ding{51}&\ding{55} &3.91\\
   		{Vocos w/o GAN}&\ding{51}&\ding{55} &3.76\\
   		{Vocos* w/o GAN}&\ding{51}&\ding{55} &3.77\\
          \hline
   		{APNet2}&\ding{51}&\ding{51} &4.00\\
   		{APNet2 w/o GAN}&\ding{51}&\ding{51} &3.89\\
   		{APNet2* w/o GAN}&\ding{51}&\ding{51} &3.95\\
          \hline
   		{FreeV}&\ding{51}&\ding{51} &3.97\\
   		{FreeV w/o GAN}&\ding{51}&\ding{51} &3.89\\
   		{FreeV* w/o GAN}&\ding{51}&\ding{51} &3.95\\
				\hline
				\hline
	\end{tabular}}}
\end{table}


\begin{figure}
    \centering
    \includegraphics[width=0.9\linewidth]{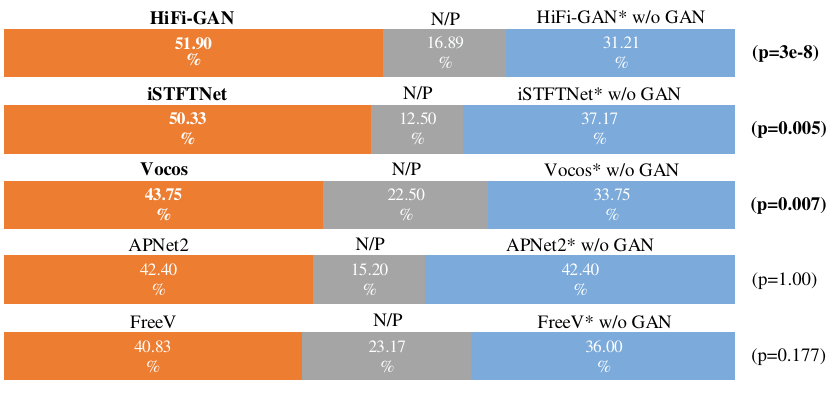}
    \caption{ Average preference scores (\%) of ABX tests on speech
quality between baseline vocoders and their GAN-free versions with our proposed improvement strategies (`* w/o GAN'), where N/P means ``no preference" and p denotes the p value of a \textit{t}-test.}
    \label{fig:2}
\end{figure}
\vspace{-1mm}
\vspace{-1mm}
\subsection{GAN is Necessary for Which Types of Vocoders?}
\label{ssec:eaq}
\vspace{-1mm}
Then, we address the question of which types of vocoders require GAN to evaluate the generalizability of the proposed improvement strategies. 
We compared HiFi-GAN, iSTFTNet, Vocos, APNet2 and FreeV with their respective GAN-free versions and GAN-free versions with the improvement strategies. 
Since we found that BigVGAN failed to converge after removing GAN, we have not included its results. 
We presented the objective UTMOS metrics in Table \ref{tab2} and ABX preference subjective test results in Figure \ref{fig:2}. 
In Table \ref{tab2}, we also labeled each vocoder based on whether it modeled the spectrum and explicitly predicted the phase, for easier classification.

As shown in Table \ref{tab2}, for all vocoders tested, removing GAN results in a significant drop in UTMOS. 
However, after applying our proposed improvement strategy, UTMOS shows a noticeable improvement.
Although HiFi-GAN* w/o GAN, iSTFTNet* w/o GAN, and Vocos w/o GAN adopted our improvement strategies, a significant gap remains compared to the original versions, as shown in both Table \ref{tab2} and Figure \ref{fig:2}. 
Although they model different targets (spectrum or waveform), they share the characteristic of not explicitly predicting the phase. 
Interestingly, for explicit-phase-prediction-based APNet2 and FreeV, despite the removal of GAN, our improvement strategy allowed them to achieve performance nearly on par with their original versions.
As shown in Figure \ref{fig:2}, no significant difference was observed between APNet2 and APNet2* w/o GAN (p=1.00), or between FreeV and FreeV* w/o GAN (p=0.177), in terms of subjective listening quality. 
This indicates that neural vocoders based on explicit phase prediction can benefit from the improvement strategies we proposed and no longer rely on GAN, whereas GAN is still necessary for other types of neural vocoders.

\vspace{-1mm}
\vspace{-1mm}
\section{Conclusion}
\vspace{-1mm}
This paper presents FreeGAN, a mel-spectrogram-based neural vocoder that does not rely on GAN training. 
It introduces improvement strategies at input, structure, and loss aspects to compensate for the performance loss caused by the absence of GAN. 
Experimental results confirm that FreeGAN can generate speech of comparable quality to other advanced GAN-based vocoders, with significant advantages in training and storage efficiency. 
Applying FreeGAN to practical tasks, e.g., speech synthesis and coding, will be our future work.

\clearpage

\bibliographystyle{IEEEtran}
\bibliography{mybib}

\begin{thebibliography}{10}
\providecommand{\url}[1]{#1}
\csname url@samestyle\endcsname
\providecommand{\newblock}{\relax}
\providecommand{\bibinfo}[2]{#2}
\providecommand{\BIBentrySTDinterwordspacing}{\spaceskip=0pt\relax}
\providecommand{\BIBentryALTinterwordstretchfactor}{4}
\providecommand{\BIBentryALTinterwordspacing}{\spaceskip=\fontdimen2\font plus
\BIBentryALTinterwordstretchfactor\fontdimen3\font minus \fontdimen4\font\relax}
\providecommand{\BIBforeignlanguage}[2]{{%
\expandafter\ifx\csname l@#1\endcsname\relax
\typeout{** WARNING: IEEEtran.bst: No hyphenation pattern has been}%
\typeout{** loaded for the language `#1'. Using the pattern for}%
\typeout{** the default language instead.}%
\else
\language=\csname l@#1\endcsname
\fi
#2}}
\providecommand{\BIBdecl}{\relax}
\BIBdecl

\bibitem{shen2018natural}
J.~Shen, R.~Pang, R.~J. Weiss, M.~Schuster, N.~Jaitly, Z.~Yang, Z.~Chen, Y.~Zhang, Y.~Wang, R.~Skerrv-Ryan \emph{et~al.}, ``Natural {TTS} synthesis by conditioning wavenet on mel spectrogram predictions,'' in \emph{Proc. ICASSP}, 2018, pp. 4779--4783.

\bibitem{du2024cosyvoice}
Z.~Du, Q.~Chen, S.~Zhang, K.~Hu, H.~Lu, Y.~Yang, H.~Hu, S.~Zheng, Y.~Gu, Z.~Ma \emph{et~al.}, ``Cosyvoice: A scalable multilingual zero-shot text-to-speech synthesizer based on supervised semantic tokens,'' \emph{arXiv preprint arXiv:2407.05407}, 2024.

\bibitem{lu2023mp}
Y.-X. Lu, Y.~Ai, and Z.-H. Ling, ``{MP-SEN}et: {A} speech enhancement model with parallel denoising of magnitude and phase spectra,'' in \emph{Proc. Interspeech}, 2023, pp. 3834--3838.

\bibitem{liu2024audiosr}
H.~Liu, K.~Chen, Q.~Tian, W.~Wang, and M.~D. Plumbley, ``{A}udio{SR}: {V}ersatile audio super-resolution at scale,'' in \emph{Proc. ICASSP}, 2024, pp. 1076--1080.

\bibitem{lu2024towards}
Y.-X. Lu, Y.~Ai, H.-P. Du, and Z.-H. Ling, ``Towards high-quality and efficient speech bandwidth extension with parallel amplitude and phase prediction,'' \emph{IEEE/ACM Transactions on Audio, Speech, and Language Processing}, pp. 1--14, 2024.

\bibitem{langman2024spectral}
R.~Langman, A.~Juki{\'c}, K.~Dhawan, N.~R. Koluguri, and B.~Ginsburg, ``Spectral codecs: Spectrogram-based audio codecs for high quality speech synthesis,'' \emph{arXiv preprint arXiv:2406.05298}, 2024.

\bibitem{stahl2024bitrate}
B.~Stahl, S.~Windtner, and A.~Sontacchi, ``A bitrate-scalable variational recurrent mel-spectrogram coder for real-time resynthesis-based speech coding,'' \emph{IEEE Access}, 2024.

\bibitem{tamamori2017speaker}
A.~Tamamori, T.~Hayashi, K.~Kobayashi, K.~Takeda, and T.~Toda, ``Speaker-dependent {WaveNet} vocoder.'' in \emph{Proc. Interspeech}, 2017, pp. 1118--1122.

\bibitem{ai2018samplernn}
Y.~Ai, H.-C. Wu, and Z.-H. Ling, ``{SampleRNN-based} neural vocoder for statistical parametric speech synthesis,'' in \emph{Proc. ICASSP}, 2018, pp. 5659--5663.

\bibitem{ai2020neural}
Y.~Ai and Z.-H. Ling, ``A neural vocoder with hierarchical generation of amplitude and phase spectra for statistical parametric speech synthesis,'' \emph{IEEE/ACM Transactions on Audio, Speech, and Language Processing}, vol.~28, pp. 839--851, 2020.

\bibitem{prenger2019waveglow}
R.~Prenger, R.~Valle, and B.~Catanzaro, ``{WaveGlow}: A flow-based generative network for speech synthesis,'' in \emph{Proc. ICASSP 2019}, pp. 3617--3621.

\bibitem{ping2020waveflow}
W.~Ping, K.~Peng, K.~Zhao, and Z.~Song, ``Wave{F}low: A compact flow-based model for raw audio,'' in \emph{Proc. ICML}, 2020, pp. 7706--7716.

\bibitem{kumar2019melgan}
K.~Kumar, R.~Zhang, C.~Fuegen, R.~Puri, Y.~Zhang, and B.~Catanzaro, ``{MelGAN}: Generative adversarial networks for conditional waveform synthesis,'' in \emph{Proc. NeurIPS}, 2019, pp. 14\,881--14\,892.

\bibitem{kong2020hifi}
J.~Kong, J.~Kim, and J.~Bae, ``{HiFi-GAN: Generative adversarial networks for efficient and high fidelity speech synthesis},'' \emph{Advances in Neural Information Processing Systems}, vol.~33, pp. 17\,022--17\,033, 2020.

\bibitem{ai2023apnet}
Y.~Ai and Z.-H. Ling, ``{APN}et: An all-frame-level neural vocoder incorporating direct prediction of amplitude and phase spectra,'' \emph{IEEE/ACM Transactions on Audio, Speech, and Language Processing}, vol.~31, pp. 2145--2157, 2023.

\bibitem{lee2022bigvgan}
S.-g. Lee, W.~Ping, B.~Ginsburg, B.~Catanzaro, and S.~Yoon, ``{BigVGAN: A universal neural vocoder with large-scale training},'' in \emph{Proc. ICLR}, 2023.

\bibitem{kaneko2022istftnet}
T.~Kaneko, K.~Tanaka, H.~Kameoka, and S.~Seki, ``{iSTFTNet: Fast and lightweight mel-spectrogram vocoder incorporating inverse short-time Fourier transform},'' in \emph{Proc. {ICASSP}}, 2022, pp. 6207--6211.

\bibitem{yamamoto2020parallel}
R.~Yamamoto, E.~Song, and J.-M. Kim, ``Parallel {WaveGAN}: A fast waveform generation model based on generative adversarial networks with multi-resolution spectrogram,'' in \emph{Proc. ICASSP}, 2020, pp. 6199--6203.

\bibitem{nguyen2024fregrad}
T.~D. Nguyen, J.-H. Kim, Y.~Jang, J.~Kim, and J.~S. Chung, ``Fregrad: Lightweight and fast frequency-aware diffusion vocoder,'' in \emph{Proc. ICASSP}, 2024, pp. 10\,736--10\,740.

\bibitem{espic2017direct}
F.~Espic, C.~Valentini-Botinhao, and S.~King, ``Direct modelling of magnitude and phase spectra for statistical parametric speech synthesis.'' in \emph{Proc. Interspeech}, 2017, pp. 1383--1387.

\bibitem{loweimi2021speech}
E.~Loweimi, Z.~Cvetkovic, P.~Bell, and S.~Renals, ``Speech acoustic modelling from raw phase spectrum,'' in \emph{Proc. {ICASSP}}, 2021, pp. 6738--6742.

\bibitem{du2023apnet2}
H.-P. Du, Y.-X. Lu, Y.~Ai, and Z.-H. Ling, ``{APNet2}: High-quality and high-efficiency neural vocoder with direct prediction of amplitude and phase spectra,'' in \emph{Proc. NCMMSC, 2023}, pp. 66--80.

\bibitem{lv2024freev}
Y.~Lv, H.~Li, Y.~Yan, J.~Liu, D.~Xie, and L.~Xie, ``Free{V}: {F}ree lunch for vocoders through pseudo inversed mel filter,'' in \emph{Proc. Interspeech}, 2024, pp. 3869--3873.

\bibitem{ai2023neural}
Y.~Ai and Z.-H. Ling, ``Neural speech phase prediction based on parallel estimation architecture and anti-wrapping losses,'' in \emph{Proc. {ICASSP}}, 2023, pp. 1--5.

\bibitem{woo2023convnext}
S.~Woo, S.~Debnath, R.~Hu, X.~Chen, Z.~Liu, I.~S. Kweon, and S.~Xie, ``{ConvNeXt v2: Co-designing and scaling convnets with masked autoencoders},'' in \emph{Proc. CVPR}, 2023, pp. 16\,133--16\,142.

\bibitem{jiang2024mdctcodec}
X.-H. Jiang, Y.~Ai, R.-C. Zheng, H.-P. Du, Y.-X. Lu, and Z.-H. Ling, ``{MDCTC}odec: A lightweight {MDCT}-based neural audio codec towards high sampling rate and low bitrate scenarios,'' in \emph{Proc. SLT}, 2024, pp. 550--557.

\bibitem{du2024bivocoder}
H.-P. Du, Y.-X. Lu, Y.~Ai, and Z.-H. Ling, ``Bivocoder: A bidirectional neural vocoder integrating feature extraction and waveform generation,'' in \emph{Proc. Interspeech}, 2024, pp. 3894--3898.

\bibitem{hendrycks2016gaussian}
D.~Hendrycks and K.~Gimpel, ``{Gaussian error linear units (GELUs)},'' \emph{arXiv preprint arXiv:1606.08415}, 2016.

\bibitem{ziyin2020neural}
L.~Ziyin, T.~Hartwig, and M.~Ueda, ``Neural networks fail to learn periodic functions and how to fix it,'' \emph{Advances in Neural Information Processing Systems}, vol.~33, pp. 1583--1594, 2020.

\bibitem{veaux2016superseded}
C.~Veaux, J.~Yamagishi, K.~MacDonald \emph{et~al.}, ``Superseded-{CSTR VCTK} corpus: English multi-speaker corpus for cstr voice cloning toolkit,'' 2016.

\bibitem{kingma2014adam}
D.~P. Kingma and J.~Ba, ``Adam: A method for stochastic optimization,'' in \emph{Proc. ICLR}, 2015.

\bibitem{siuzdakvocos}
H.~Siuzdak, ``Vocos: Closing the gap between time-domain and fourier-based neural vocoders for high-quality audio synthesis,'' in \emph{Proc. ICLR}, 2024.

\bibitem{saeki2022utmos}
T.~Saeki, D.~Xin, W.~Nakata, T.~Koriyama, S.~Takamichi, and H.~Saruwatari, ``{UTMOS: Utokyo-sarulab system for voiceMOS Challenge 2022},'' in \emph{Proc. Interspeech}, 2022, pp. 4521--4525.

\end{thebibliography}

\end{document}